\documentclass{article}
\usepackage{spconf,amsmath,graphicx}
\usepackage{cite}
\usepackage{amssymb,amsfonts,amsthm,bm}
\usepackage[usenames,dvipsnames]{xcolor}

\usepackage{booktabs}
\usepackage{graphicx}
\usepackage{multirow}
\usepackage{stfloats}
\usepackage{array} 
\usepackage{caption}
\usepackage{algorithm}
\usepackage{algorithmic}

\definecolor{mygreen}{rgb}{0,0.5,0}
\definecolor{darkblue}{RGB}{0,0,150}
\title{Deep Mamba Multi-modal Learning}
\name{Jian Zhu$^{1}$, Xin Zou$^{2}$, Yu Cui$^{1}$, Zhangmin Huang$^{1}$, Chenshu Hu$^{1}$, Bo Lyu$^{1}$} 
\address{$^{1}$ Zhejiang Lab, China \{qijian.zhu, yu.cui, zmhuang, hucs\_frog, bo.lyu\}@zhejianglab.edu.cn\\	$^{2}$  China University of Geosciences, China \{zouxin\}@cug.edu.cn}
\vspace{-0.35cm}
\begin{document}
\ninept
\maketitle
\ninept
\begin{abstract}
Inspired by the excellent performance of Mamba networks, we propose a novel Deep Mamba Multi-modal Learning (DMML). It can be used to achieve the fusion of multi-modal features. We apply DMML to the field of multimedia retrieval and propose an innovative Deep Mamba Multi-modal Hashing (DMMH) method. It combines the advantages of algorithm accuracy and inference speed. We validated the effectiveness of DMMH on three public datasets and achieved state-of-the-art results.
\end{abstract}

\section{Introduction}
% Multi-modal hashing is one of the important technologies in the field of multimedia retrieval. It is the fusion of multi-modal heterogeneous data to generate hash codes.

\begin{figure}
	\centering
	\includegraphics[width=9cm]{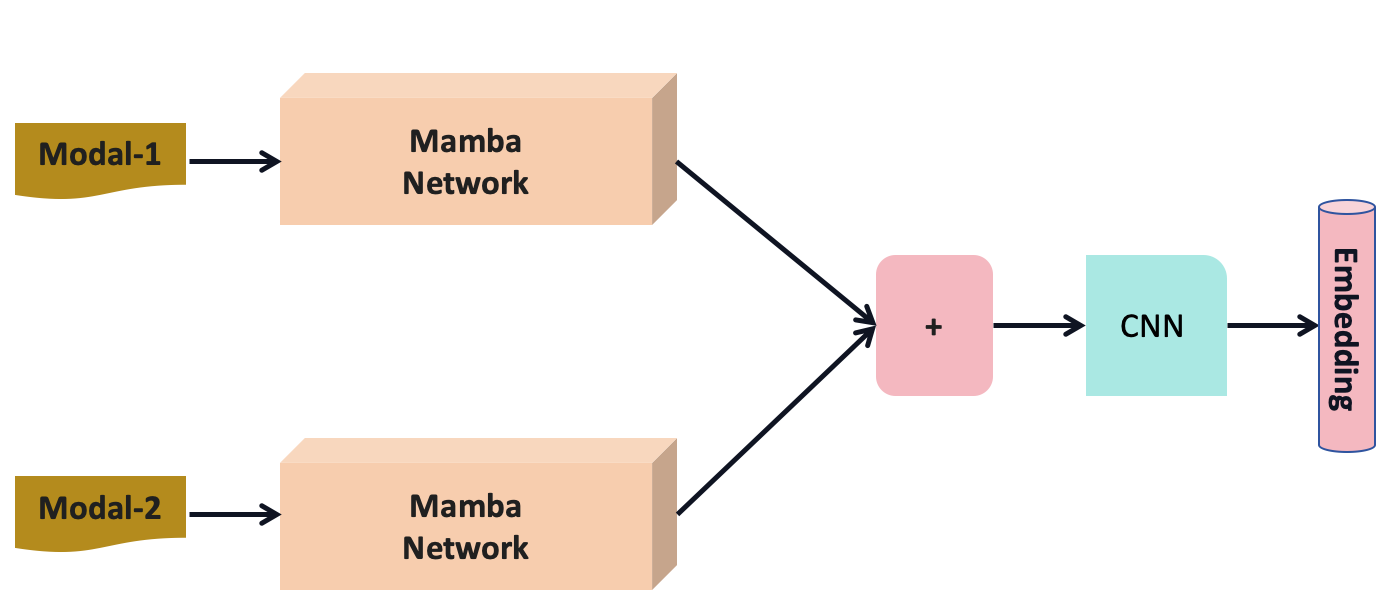}
	\caption{Deep Mamba Multi-modal Learning}
	\label{fig:01}
\end{figure}

% The current multi-modal hashing methods have the problem of low retrieval accuracy. The reason is that the backbone networks lack good feature expression capability. For instance, Flexible Multi-modal Hashing (FDH) \cite{zhu:52} and Bit-aware Semantic Transformer Hashing (BSTH) \cite{tan:61} hire a VGG net \cite{simonyan:51} for the image modal. And these methods use a Bag-of-Words model \cite{zhang:53} for the text modal. They are outdated for feature extraction, thus, an update in feature extraction methods is necessary. The fact above results in a degradation of the overall retrieval accuracy for the current multi-modal hashing method.
We propose a new Deep Mamba Multi-modal Learning (DMML) method. As shown in Fig. \ref{fig:01}, it is made up of Mamba networks and CNNs. Firstly, it uses a Mamba network to enhance the semantic features of a single modality. Secondly, we adopt an additive method to achieve a simple fusion of multi-modal features. Finally, it uses CNNs for the deep fusion of multi-modal features to enhance semantic expression ability.

% In recent years, multi-modal large-scale models have achieved great success. Because these models are trained on large-scale data, they have stronger semantic expression ability. Contrastive Language-Image Pre-training(Mamba Network) \cite{radford:63} is one of the most representative multi-modal models. However, The application of a multi-modal large model in multi-modal retrieval has not been studied. For the first time, we investigate how Mamba Network affects the retrieval efficiency of multi-view hashing. As shown in Fig. \ref{fig:01}, it is pre-trained by contrastive learning on large-scale image text data pairs. It has shown exceptional zero-shot or few-shot learning abilities as well as excellent semantic understanding capabilities. The multi-modal field has been greatly changed by Mamba Network, and more people are beginning to acknowledge that it is superior at multi-modal tasks. Although Mamba Network has undergone multiple successful trials, a thorough analysis of its effects and performance on multi-modal hashing retrieval has not yet been conducted.

% We perform an in-depth study in this work to examine the potential of the Mamba Network on retrieval from multi-modal hashing. We use the Mamba Network model to extract text and image features. The extracted modal feature data through the Mamba Network model performs better. It can significantly improve the retrieval performance of multi-modal hashing methods. Compared with the latest state-of-the-art method, the DMMH proposed by us has a maximum improvement of $3.38\%$.
Based on DMML, we propose a novel Deep Mamba Multi-modal Hashing (DMMH) method. Our DMMH method is validated for effectiveness on three datasets and achieves state-of-the-art results in multimedia retrieval.

Our contributions are as follows:
\begin{itemize}
\item We are using the Mamba network for the first time in multimedia retrieval.
\item We present a novel Deep Mamba Multi-modal Learning (DMML) method to fuse heterogeneous data.
\item We propose a new Deep Mamba Multi-modal Hashing (DMMH), which achieves the state-of-the-art result in multimedia retrieval.
\end{itemize}

\begin{figure*}
	\centering
	\includegraphics[width=18cm]{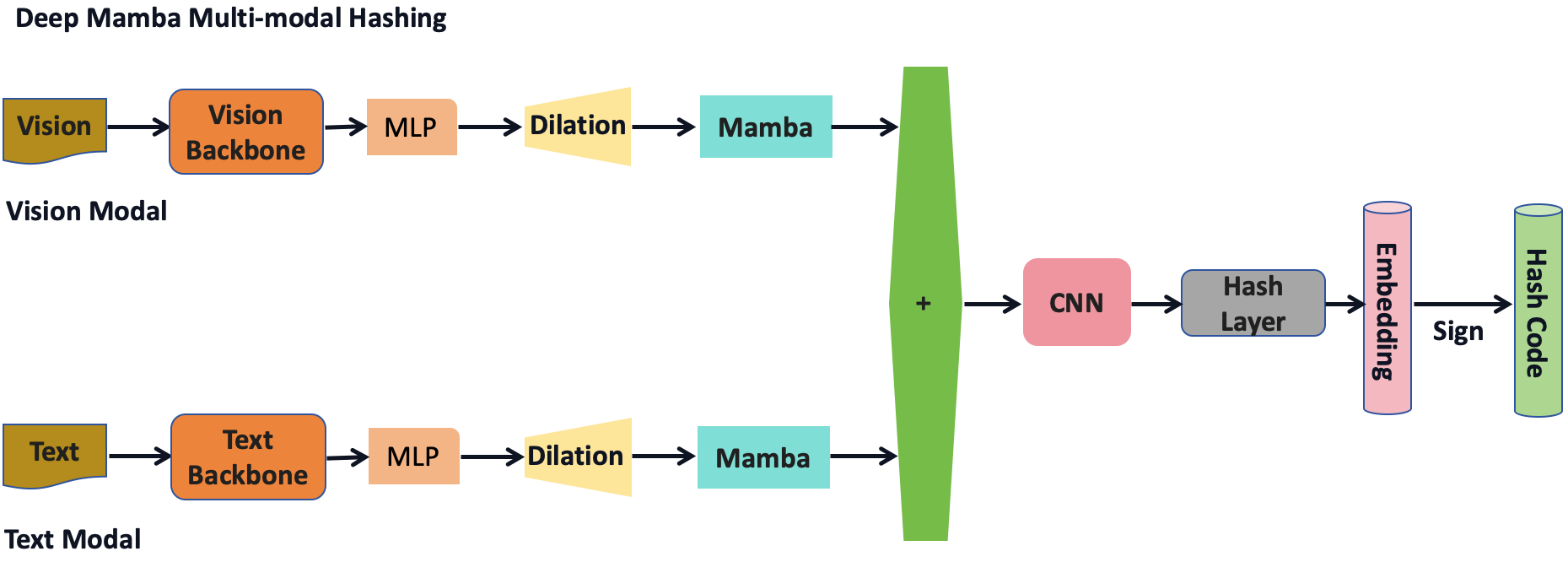}
	\caption{Deep Mamba Multi-modal Hashing}
	\label{fig:02}
\end{figure*}

\section{Methodology}
Deep multi-modal hashing network aims to convert multi-modal data into hash code. As shown in Fig. \ref{fig:02}, DMMH consists of backbones, MLPs, dilation networks, Mamba networks, CNNs, and a hash layer. These modules are described in detail below.
\begin{enumerate}
\item \textbf{Vision Backbone:} VGGNet is used to generate visual features. 

\item \textbf{Text Backbone:}  BoW model is utilized to produce text features. 

\item \textbf{MLPs:} We use MLPs networks for feature normalization. It achieves alignment of multi-modal feature dimensions.

\item \textbf{Dilation:} Dilation networks mainly inflate single modal features and convert them into sequence embeddings.

\item \textbf{Mamba Network:} We use Mamba networks for fine-grained semantic mining.

\item \textbf{CNNs:} We use CNNs for deep fusion of multi-modal features.

\item \textbf{Hash Layer:} The hash layer converts the learned embeddings into binary hash codes.
\end{enumerate}

\begin{table*}
\centering
\caption{An statistical overview of three public datasets.}
\begin{tabular}{llllllll}
\toprule[1pt]
Dataset   & Training Size & Retrieval Size & Query Size & Categories&Vision Embedding & Text Embedding \\ \midrule[0.8pt]
MIR-Flickr25K & 5000  & 17772 & 2243    & 24&4096-D & 1386-D \\
NUS-WIDE & 21000  & 193749 & 2085    & 21&4096-D &1000-D\\
MS COCO & 18000  & 82783 & 5981    & 80&4096-D &2000-D\\
\bottomrule[1pt]
\end{tabular}
\label{Tab:01}
\end{table*}

\begin{table*}
	\setlength{\tabcolsep}{2pt}
	\centering
	\caption{The comparable mAP results on MIR-Flickr25K, NUS-WIDE, and MS COCO. The best results are bolded, and the second-best results are underlined. The * indicates that the results of our method on this dataset are statistical significance.}
	\resizebox{\textwidth}{!}{\begin{tabular}{llllllllllllll}
		\toprule[1pt]
		\multicolumn{1}{c}{\multirow{2}{*}{Method}} & \multicolumn{1}{c}{\multirow{2}{*}{Ref.}} & \multicolumn{4}{c}{MIR-Flickr25K*}    & \multicolumn{4}{c}{NUS-WIDE*}      & \multicolumn{4}{c}{MS   COCO*}       \\  \cmidrule(r){3-6}  \cmidrule(r){7-10}  \cmidrule(r){11-14}
		\multicolumn{1}{c}{}                         & \multicolumn{1}{c}{}                      & 16 bits & 32 bits & 64 bits & 128 bits & 16 bits & 32 bits & 64 bits & 128 bits & 16 bits & 32 bits & 64 bits & 128 bits \\ \midrule[0.8pt]
		MFH                                          & TMM13                                     & 0.5795 & 0.5824 & 0.5831 & 0.5836  & 0.3603 & 0.3611 & 0.3625 & 0.3629  & 0.3948 & 0.3699 & 0.3960  & 0.3980   \\
		MAH                                          & TIP15                                     & 0.6488 & 0.6649 & 0.6990  & 0.7114  & 0.4633 & 0.4945 & 0.5381 & 0.5476  & 0.3967 & 0.3943 & 0.3966 & 0.3988  \\
		MVLH                                         & MM15                                      & 0.6541 & 0.6421 & 0.6044 & 0.5982  & 0.4182 & 0.4092 & 0.3789 & 0.3897  & 0.3993 & 0.4012 & 0.4065 & 0.4099  \\
		MvDH                                         & TIST18                                    & 0.6828 & 0.7210  & 0.7344 & 0.7527  & 0.4947 & 0.5661 & 0.5789 & 0.6122  & 0.3978 & 0.3966 & 0.3977 & 0.3998  \\ \midrule[0.8pt]
		MFKH                                         & MM12                                      & 0.6369 & 0.6128 & 0.5985 & 0.5807  & 0.4768 & 0.4359 & 0.4342 & 0.3956  & 0.4216 & 0.4211 & 0.4230  & 0.4229  \\
		DMVH                                         & ICMR17                                    & 0.7231 & 0.7326 & 0.7495 & 0.7641  & 0.5676 & 0.5883 & 0.6902 & 0.6279  & 0.4123 & 0.4288 & 0.4355 & 0.4563  \\
		FOMH                                         & MM19                                      & 0.7557 & 0.7632 & 0.7564 & 0.7705  & 0.6329 & 0.6456 & 0.6678 & 0.6791  & 0.5008 & 0.5148 & 0.5172 & 0.5294  \\
		FDMH                                         & NPL20                                     & 0.7802 & 0.7963 & 0.8094 & 0.8181  & 0.6575 & 0.6665 & 0.6712 & 0.6823  & 0.5404 & 0.5485 & 0.5600   & 0.5674  \\
		DCMVH                                        & TIP20                                     & 0.8097 & 0.8279 & 0.8354 & 0.8467  & 0.6509 & 0.6625 & 0.6905 & 0.7023  & 0.5387 & 0.5427 & 0.5490  & 0.5576  \\
		SAPMH                                        & TMM21                                      & 0.7657 & 0.8098 & 0.8188 & 0.8191  & 0.6503 & 0.6703 & 0.6898 & 0.6901  & 0.5467 & 0.5502 & 0.5563 & 0.5672  \\
		FGCMH &MM21 & \underline{0.8173} & \underline{0.8358} & 0.8377 & \underline{0.8606}  & 0.6677 & 0.6874 & 0.6936 & 0.7011  & 0.5641 & 0.5273 & 0.5797 & 0.5862 \\
  BSTH & SIGIR22  &0.8145&0.8340&\underline{0.8482}&0.8571                                     & \underline{0.6990} & \underline{0.7340} & \underline{0.7505} & \underline{0.7704}  & \underline{0.5831} & \underline{0.6245} & \underline{0.6459} & \underline{0.6654}  \\\midrule[0.8pt]
DMMH                                         & ours                                         &\textbf{0.8319} & \textbf{0.8523} & \textbf{0.8694} & \textbf{0.8788} & \textbf{0.7229} & \textbf{0.7424} & \textbf{0.7661} & \textbf{0.7806}  &\textbf{0.5869} & \textbf{0.6299} & \textbf{0.6477} & \textbf{0.6715} \\
		\bottomrule[1pt]
	\end{tabular}}

	\label{Tab:02}
\end{table*}

\section{Experiments}
\subsection{Evaluation Datasets and Metrics}
We assess the effectiveness of the proposed DMMH method in multimedia retrieval tasks. We use three public datasets: MIR-Flickr25K \cite{huiskes:20}, NUS-WIDE \cite{chua:22}, and MS COCO \cite{lin:21}. We use mean Average Precision (mAP) as the analysis metric. An overview of the dataset statistics utilized in the studies is given in Table \ref{Tab:01}.

\subsection{Baseline Methods}
To determine the retrieval metric, we compare the proposed DMMH method with thirteen multi-view hashing methods, including four unsupervised methods (e.g., Multiple Feature Hashing (MFH) \cite{song:9}, Multi-view Alignment Hashing (MAH) \cite{liu:10}, Multi-view Latent Hashing (MVLH) \cite{shen:6}, and Multi-view Discrete Hashing (MvDH) \cite{shen:11}) and nine supervised methods (e.g., Multiple Feature Kernel Hashing (MFKH) \cite{liu:7}, Discrete Multi-view Hashing (DMVH) \cite{yang:12}, Flexible Discrete Multi-view Hashing (FDMH) \cite{liu:23}, Flexible Online Multi-modal Hashing (FOMH) \cite{lu:24}, Deep Collaborative Multi-View Hashing (DCMVH) \cite{zhu:17}, Supervised Adaptive Partial Multi-view Hashing (SAPMH) \cite{zheng:25}, Flexible Graph Convolutional Multi-modal Hashing (FGCMH) \cite{lu:18}, Bit-aware Semantic Transformer Hashing (BSTH) \cite{tan:61}, ACMVH \cite{zhu:89}, FMMVH \cite{zhu:90} and Deep Metric Multi-View Hashing (DMMVH) \cite{zhu:62}).

\subsection{Analysis of Experimental Results}
% \usepackage{multirow}

Table \ref{Tab:02} illustrates the results of the mAP. The results show that the DMMH performs significantly better than any of the examined multi-view hashing methods. In particular, our technique delivers an average mAP improvement of $2.00\%$, $1.43\%$, and $4.80\%$ on the MIR-Flickr25K, NUS-WIDE, and MS COCO datasets, respectively, when compared to the state-of-the-art multi-modal hashing method DMMVH \cite{zhu:62}. There are two primary reasons for these better results:

\begin{itemize}
\item The Mamba network has excellent representation ability. 
\item Deep Mamba Multi-modal Learning (DMML) has excellent fusion ability of heterogeneous data. 
\end{itemize}

\section{Conclusion and Future Work}
To fuse multiple source data, we present a new multi-modal learning method termed DMML. On the basis of DMML, we propose an innovative Deep Mamba Multi-modal Hashing (DMMH) method. Experiments demonstrate that the proposed DMMH achieves state-of-the-art results in multimedia retrieval tasks. In the future, we will work out more application problems of the Mamba network in multimedia retrieval.

\vspace{-9pt}
  
\ninept 
\bibliographystyle{IEEEtran}
\bibliography{IEEEabrv,myabrv_new,my_reference}
\end{document}